\documentclass[conference,10pt]{IEEEtran}

\ifCLASSINFOpdf
\usepackage[pdftex]{graphicx}
\graphicspath{{../pdf/}{../jpeg/}}
\DeclareGraphicsExtensions{.pdf,.jpeg,.png}
\else
\usepackage[dvips]{graphicx}
\graphicspath{{../eps/}}
\DeclareGraphicsExtensions{.eps}
\fi
\usepackage{cite}
\usepackage{algorithmic}
\usepackage{amssymb}
\usepackage{textcomp}

\usepackage{amsmath}
\usepackage{amsfonts}
\usepackage{flushend}
\usepackage{amssymb}
\usepackage{makecell}
\usepackage{booktabs}

\ifCLASSOPTIONcompsoc

\usepackage[caption=false,font=normalsize,labelfont=sf,textfont=sf]{subfig}
\else
\usepackage[caption=false,font=footnotesize]{subfig}
\fi
\usepackage{graphicx}

\usepackage{epsfig}
\usepackage{epstopdf}
\usepackage{color}

\let\emph\textit
\usepackage{fixltx2e}

\usepackage{multicol}  
\usepackage{multirow}

\def\BibTeX{{\rm B\kern-.05em{\sc i\kern-.025em b}\kern-.08em
		T\kern-.1667em\lower.7ex\hbox{E}\kern-.125emX}}

\begin{document}
	
\title{Analysis of Near-Field Effects, Spatial Non-Stationary Characteristics Based on 11-15 GHz Channel Measurement in Indoor Scenario}

%
%
%

\author{Haiyang~Miao, Pan~Tang, Weirang~Zuo, Qi~Wei, Lei Tian, Jianhua~Zhang \\

	\IEEEauthorblockA{State Key Lab of Networking and Switching Technology,
		\\
		Beijing University of Posts and Telecommunications, Beijing, China\\
		$\rm \left \{hymiao; tangpan27; jhzhang\right\}$@bupt.edu.cn}

	}


\markboth{Journal of \LaTeX\ Class Files,~Vol.~14, No.~8, August~2021}%
{Shell \MakeLowercase{\textit{et al.}}: A Sample Article Using IEEEtran.cls for IEEE Journals}

%

\maketitle

\begin{abstract}
	
	 In the sixth-generation (6G), with the further expansion of array element number and frequency bands, the wireless communications are expected to operate in the near-field region. The near-field radio communications (NFRC) will become crucial in 6G communication systems. The new mid-band (6-24 GHz) is the 6G potential candidate spectrum. In this paper, we will investigate the channel measurements and characteristics for the emerging NFRC. First, the near-field spherical-wave signal model is derived in detail, and the stationary interval (SI) division method is discussed based on the channel statistical properties. Then, the influence of line-of-sight (LOS) and obstructed-LOS (OLOS) environments on the near-field effects and spatial non-stationary (SnS) characteristic are explored based on the near-field channel measurements at 11-15 GHz band. We hope that this work will give some reference to the NFRC research.
	 
\end{abstract}


\IEEEpeerreviewmaketitle

\section{Introduction}

\par At present, the researches on sixth-generation (6G) has been carried out on a global scale [1-2]. In June 2023, the 44th meeting of International Telecommunication Union-Radiocommunication Sector (ITU-R) WP5D described the 6G overall objectives and trends [3]. For fulfilling these goals and trends, tremendous research efforts are being made to develop new wireless technologies to meet the key performance indicators of 6G, which are superior to those of fifth-generation (5G). For instance, thanks to the enormous spatial multiplexing and beamforming gain, the extreme multiple-input-multiple-output (E-MIMO) is expected to accomplish a 10-fold increase in the spectral efficiency for 6G [1]. Meanwhile, the unique channel characteristics and models need to be explored. In December 2023, the 3GPP Rel-19 has determined the need to study the near-field channel for 7-24 GHz.

\par At present, some research has been carried out to explore the channel for large scale antenna array. In [4], the received power and average power delay distribution (APDP) were studied in 2.6 GHz band, and the non-stationarity was analyzed in the spatial domain. In [5], two array configurations (including virtual ULA and UCA) were used to conduct channel measurement in an outdoor stadium scenario in the 1.4725 GHz band, and the angular power spectrum along the virtual linear array was analyzed. In [6], to gain further insight into the massive MIMO channel, the channel information was extracted from measurement data. In [7], the delay spread characteristic of a massive MIMO system was analyzed at 3.5 GHz in outdoor-to-indoor (O2I) scenario. In [8], the virtual antenna UCA at Tx and the dual-polarized 64-port uniform circular patch array at Rx were used for channel measurement in the urban scenario in 2.53 GHz band. In [9], the massive MIMO channel measurements were studied in terahertz band. In [10], we have proposed a channel modeling method based on the capture and observation of multipath propagation mechanisms at mmWave band. In [11], an indoor massive virtual array channel measurement was presented from 26 to 30 GHz. In [12], the evolution of cluster number with virtual array was given in the 3.5 GHz band under line-of-sight (LOS) and non-line-of-sight (NLOS) conditions respectively, which showed the cluster appearance and disappearance.

\par Most channel measurements were carried out at 1.4725 GHz, 2.53 GHz, 2.6 GHz, 3.5 GHz, and high-bands. Currently, the new mid-band (6-24 GHz) has attracted extensive attention from academia and industry, becoming the potential band for 6G [13]. Some companies and researchers expressed their considerations for 6G spectrum. Ericsson thinks the 7-15 GHz range can be exploited for 6G by deploying advanced sharing mechanisms. In [14], the system performances at the new mid-band were discussed by comparing with sub-6 GHz and mmWave. Notably, in November 2023, WRC-23 proposed to identify the mid-band for mobile in every ITU Region [15].

\par In this paper, the NFRC research work is mainly aimed at the theoretical derivation, measurement and characterization for the near-field channel. The main contributions and novelties of this paper are summarized as follows:

\begin{itemize}

\item The near-field spherical-wave signal model is analyzed and discussed theoretically. The concept of multiplanar wave (MW) model is proposed based on the division of stationary interval (SI). The new criterion is introduced to discuss the partition method of SI with the adaptive division method.

\item Based on the near-field channel measurements in the new mid-band, the differences of channel characteristics and SI in the different environments are explored for the near-field effect in the LOS and obstructed-LOS (OLOS) environments. Especially, the spatial non-stationarity on antenna array are validated by investigating the temporal-spatial channel characteristics including power, phase, delay spread, etc.

\end{itemize}

\par The rest of this paper is organized as follows. In Section II, the near-field signal model is deduced theoretically, and the division methods of SI are analyzed. In Section III, the verification of near-field channel characteristics are explored under LOS/OLOS environments at the new mid-band. Finally, we conclude this paper in Section IV.


\section{ Analysis of Spherical-wave Model and Spatial Non-stationary for the NFRC}

\par With the increase of MIMO scale, near-field effects and SnS characteristics are more likely to appear. The large physical size of the array will lead to the different large/small scale fading characteristics of different region on the arrays. The SnS characteristics refer to the different channel environments seen, so the inevitable phenomenon is the birth-death process of the ray/cluster. It can be found that the SnS phenomena includes the channel characteristics change caused by occlusion, and the channel characteristics will also change greatly along the elements under the LOS condition in near-field region. However, the large/small scale fading characteristics of all elements are the same within the stationary interval (SI), which can be assumed that the far-field condition is approximately satisfied. In the near-field region, there is an important issue: How to characterize the spherical wave signal model in near-field region?

\begin{figure}[htbp]
	\setlength{\abovecaptionskip}{0.1 cm}
	\centering
	\includegraphics[width=0.42\textwidth]{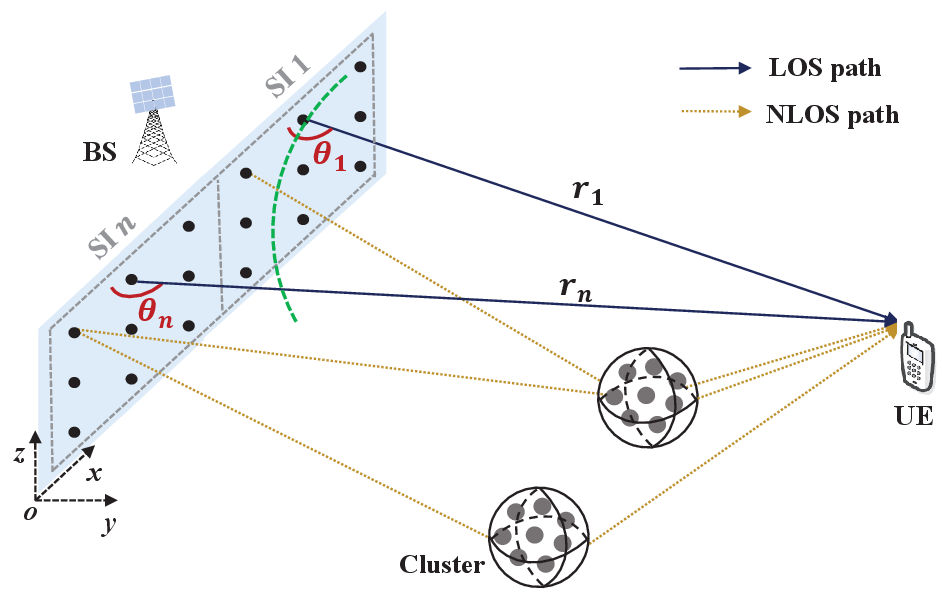}
	\caption{Near-field multipath channel propagation model.}
	\label{fig:Figure1}
\end{figure}

\subsection{Spherical-Wave Model}

\par For channel modeling, compared with distance estimation, angle estimation has more high-precision SAGE algorithms [6]. Therefore, we consider introducing the angle to model the spherical wave signal. The BS on the base station side is a large-scale array in Fig. 1, and we take the first hop cluster or UE on the downlink as an example. In the SI, the signal propagates in the form of planar wave, so the multipath information (including angle, power, etc.) can be obtained based on estimation in the same SI. The spherical-wave model needs to be established between different SIs, thus reducing the influence of the complexity brought by spherical wave.

\par The angle of path is $\theta_1$ and the distance of path is $r_1$ between UE and $SI$1 subarray as a reference subarray. The angle of path is $\theta_n$ and the distance of path is $r_n$ between UE and $SI$n subarray, and $\delta_n$ is the wave path difference due to spherical wave. In Fig. 1, the $r_1$ can be defined as

\begin{equation}
\begin{split}
r_1 & = \dfrac{(n-1)d}{sin(\theta_n-\theta_1)}sin(\pi-\theta_n)=\dfrac{(n-1)d}{sin(\theta_n-\theta_1)}sin(\theta_n).
\end{split} 
\end{equation}

Also, the $r_n$ can be defined as

\begin{equation}
\begin{split}
	r_n & =\dfrac{(n-1)d}{sin(\theta_n-\theta_1)}sin(\theta_1).
\end{split} 
\end{equation}

Then, the path difference $\delta_n$ can be expressed as

\begin{equation}
\begin{split}
\delta_n & = r_n-r_1 = \dfrac{(n-1)d}{sin(\theta_n-\theta_1)}(sin(\theta_1)-sin(\theta_n)).
\end{split} 
\end{equation}

Therefore, the expression of the phase of the near-field is derived as

\begin{equation}
\begin{split}
\varphi_n & = \dfrac{2\pi \delta_n}{\lambda} = \dfrac{2\pi}{\lambda}(n-1)d\dfrac{sin(\theta_1)-sin(\theta_n)}{sin(\theta_n-\theta_1)}.
\end{split} 
\end{equation}

In the far-field, the signal propagates in the form of the planar wave, and the $\theta_n$ of any element reaching the array surface is equal. In the above phase formula, the phase result of the far field can be obtained in the limit case, that is, $\theta_n=\theta_1$. $\dfrac{2\pi}{\lambda}(n-1)d$ is a parameter related to the number of elements and the spacing. The following is the limit value derivation of the angular domain, take the derivative of $\theta_n$ as

\begin{equation}
\begin{split}
\lim_{\theta_n=\theta_1} \dfrac{sin(\theta_1)-sin(\theta_n)}{sin(\theta_n-\theta_1)}& = \lim_{\theta_n=\theta_1} \dfrac{-cos(\theta_n)}{cos(\theta_n-\theta_1)}=-cos(\theta_1).
\end{split} 
\end{equation}

Thus, the phase absolute value of the far-field can be expressed as

\begin{equation}
\begin{split}
\varphi_n & = \dfrac{2\pi d(n-1)cos(\theta_1)}{\lambda}. 
\end{split} 
\end{equation}

\par It can be seen from the above model proof that there is a transformation relationship between near and far fields. The far field is a limiting case of the near field, or a special case, which is solved by the limit value.

\par Then the accuracy of the model is proved in a Wireless Insite simulation scenario where the originating 256-element array antenna, as shown in Fig. 2(a). Fig. 2(b) shows the phase simulation results and the theoretical results using the spherical wave signal model.

\begin{figure}[htbp]
	\xdef\xfigwd{\columnwidth}
	\setlength{\abovecaptionskip}{0.1 cm}
	\centering
	\begin{tabular}{cc}
		\includegraphics[width=4.2cm,height=3cm]{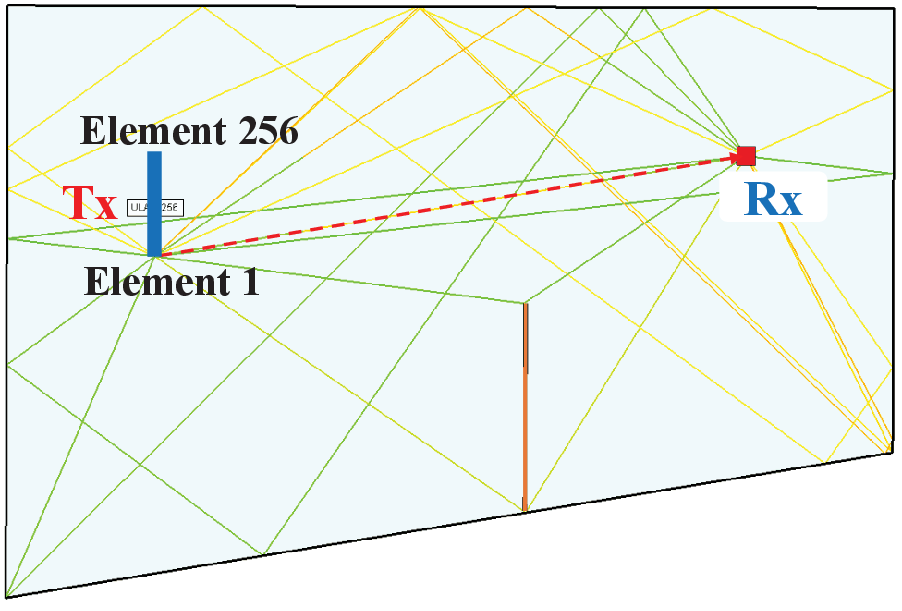} \hspace{-5mm} &\includegraphics[width=4.5cm,height=3.2cm]{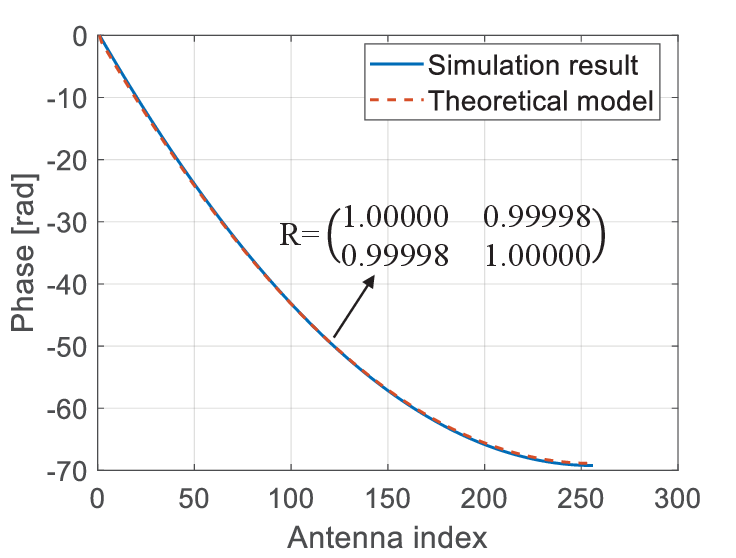}\\
		{\footnotesize\sf (a)} &
		{\footnotesize\sf (b)} \\
	\end{tabular}
	\caption{Phase verification. (a) Simulation scenario. (b) Simulation results and theoretical model.}
	\label{fig:Figure2}
\end{figure}

\par In Fig. 2(b), the phase obtained by simulation is very similar to the established signal model. Using correlation to judge the two groups of numbers, it can be observed that the correlation between the two groups of results is very high, indicating that the accuracy of theoretical model is ideal.





\subsection{Spatial Non-Stationary}

\par In this paper, the subarray of the antenna array is defined as the SI, where the large/small scale fading characteristics are stationary. At present, the division method of SI is usually uniform division based on the experience, but does not consider environmental factors. In the actual environment, the channel is dynamic due to the randomness of the scattering environment, that is, the channel matrix presents a nonlinear structure, and the SIs of antenna array are more likely to exhibit irregularity. Therefore, this paper proposes two potential adaptive division methods.

\par \textbf{\emph{Based on channel correlation}}: Similarity measurement is a key task in the process of stationary interval division of array dimension. Based on the distance criterion, the correlation matrix distance is introduced to measure, whose measurement quality directly affects the accuracy of modeling. Correlation matrix distance measures the difference between two correlation matrices, which can be understood as the orthogonality between matrices. The channel correlation matrix is calculated, and the correlation between different sub-channels is determined by calculating Pearson correlation coefficient. Using the channel correlation matrix, the correlation matrix distance $D$ is introduced to measure the similarity between two subintervals.

\begin{equation}
\begin{split}
D(R_1,R_2) & = 1-\dfrac{tr({R_1,R_2})}{\parallel R_1\parallel_{F} \parallel R_2\parallel_{F}},
\end{split} 
\end{equation}
where $R_1$ and $R_2$ represent the autocorrelation forms of subarray. $tr$ is the trace of a matrix. $\parallel \cdot \parallel_{F}$ is Frobenius norm (F-norm). The value of $D$ approaching 1 indicates that the matrices are independent, that is, the difference is maximized.

\par \textbf{\emph{Based on channel characteristic}}: Instead of phase variations as for Rayleigh distance, the uniform-power distance concerns the signal amplitude/power variations across array elements, where the power ratio between the weakest and strongest array element is no smaller than a certain threshold. We find that the angle on the array is constant in the far-field, which can better describe the difference between the near-field and the far-field. In our work, we can extract the angle of paths from all receiving ends to all transmitting antenna array elements.

\par To analyze the fluctuation of channel characteristic parameters $s \in [DS, AS, SF, K, \mathbf{\Omega}_{1},\mathbf{\Omega}_{2},\tau, P]$ on the array domain, some modeling methods can be used for the single parameter, such as the variance, range, coefficient of variation, etc. The slope $k$ of the model curve that the parameters $s$ on the array can be used for measurement.

\begin{equation}
\begin{split}
k & \triangleq \lim_{\Delta n \to 0} \dfrac{s(n+\Delta n)-s(n)}{\Delta n},
\end{split} 
\end{equation}
where DS, AS, SF, K and P are delay spread, angular spread, shadow fading, Ricean K-factor and power, respectively. $n$ is the index of the array elements.

\begin{figure}[htbp]
	\xdef\xfigwd{\columnwidth}
	\setlength{\abovecaptionskip}{0.1 cm}
	\centering
	\begin{tabular}{cc}
		\includegraphics[width=4.2cm,height=3.5cm]{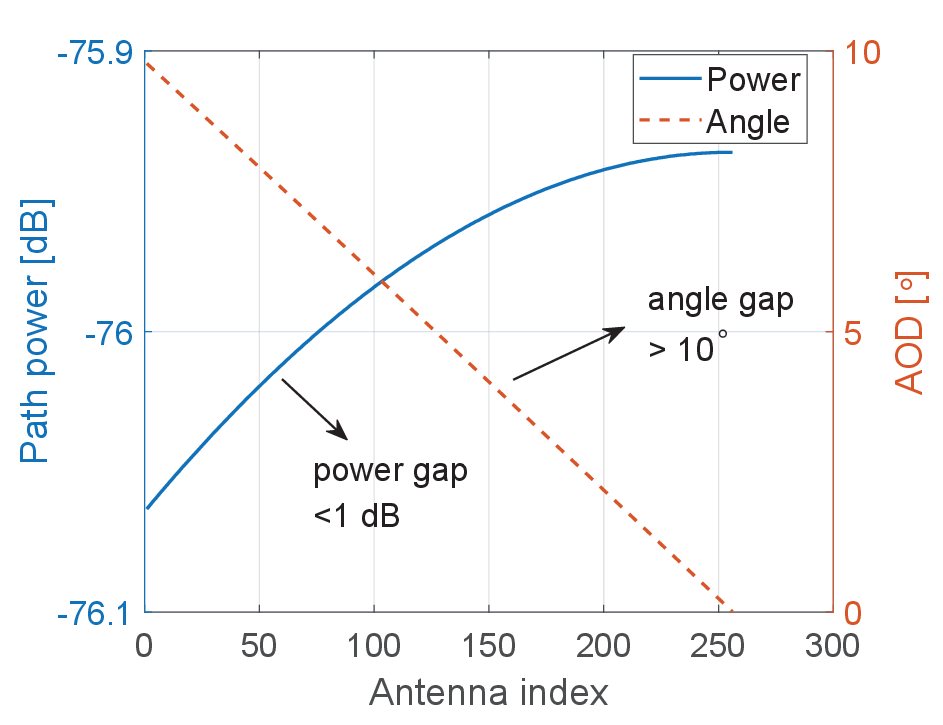} \hspace{-5mm}
		& \includegraphics[width=4.2cm,height=3.5cm]{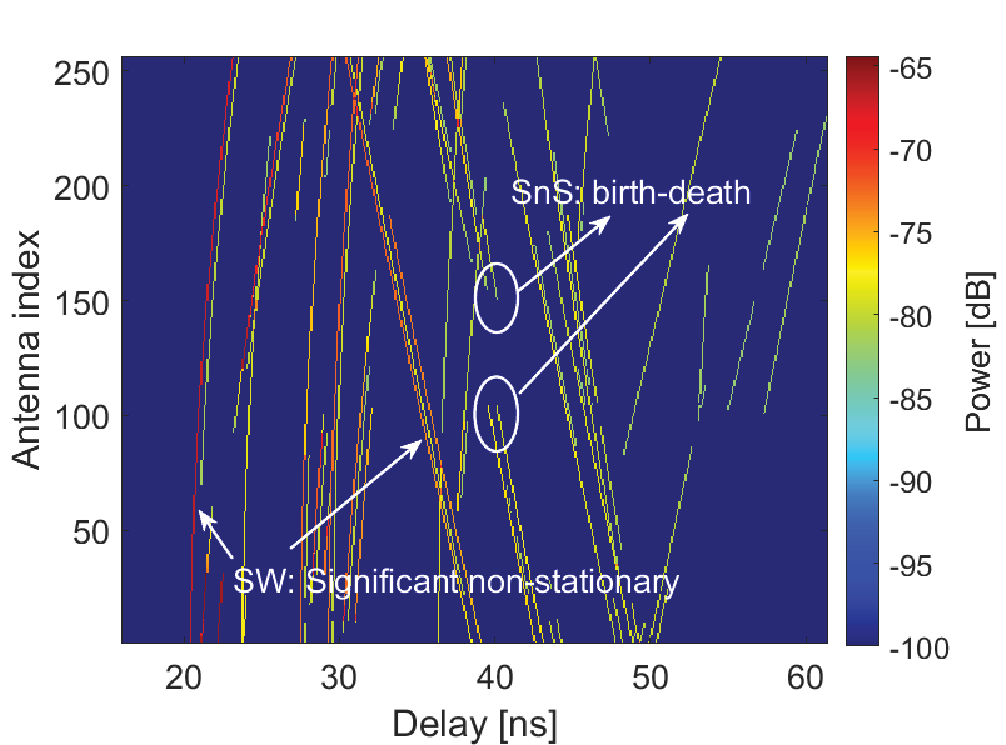}\\
		{\footnotesize\sf (a)} &	{\footnotesize\sf (b)} \\		
	\end{tabular}
	\caption{The near-field channel. (a) Power and AOD of LOS path. (b) Spherical wave effect and non-stationary.}
	\label{fig:Figure3}
\end{figure}

\par  Fig. 3(a) shows that the maximum power difference is only 0.12 dB across array elements, and the change of power is not obvious in the array domain. The main reason is that the power is a large-scale parameter. However, the largest difference is obvious for angle with 10$^\circ$. The LOS power on the array element follows the quadratic polynomial model. The LOS angle presents a linear change with the change of the array elements, indicating the occurrence of SnS phenomenon.

\par In [16], we proposed a SnS simulation framework from the cluter visibility probability. Considering the characteristics distribution on the array, we continue the consideration of simulation framework evolution in this section, to generate non-stationarty for the base station array, the array is divided into subarrays, i.e., stationary intervals. Within the SI, the scatterer environment seen by all elements is the same. Therefore, it can be considered that the traditional far-field modeling method can be adopted in the stationary intervals. Modeling method based on planar wave under far-field hypothesis. On the one hand, it is obviously unreasonable to use plane wave modeling for the entire array. On the other hand, the spherical wave modeling is adopted for each array element, which increases the complexity of the channel model. As shown in Fig. 4, the antenna array is divided into three SIs, where the far-field hypothesis can be regarded as satisfied within the SIs. The channel characteristic model is modified between SIs based on near-field effect.

\begin{figure}[htbp]
	\setlength{\abovecaptionskip}{0.1 cm}
	\centering
	\includegraphics[width=0.42\textwidth]{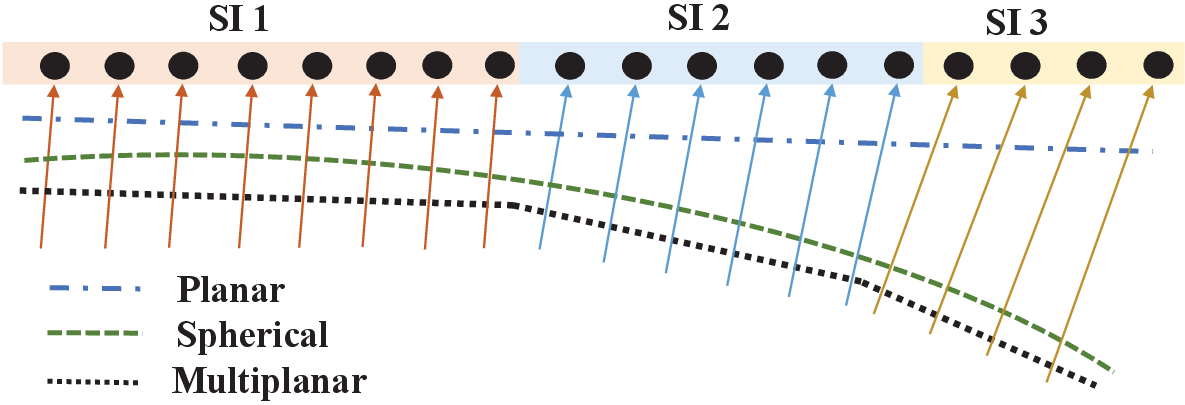}
	\caption{Division of the base station array into sub-arrays according to the stationarity interval. Wavefront: Planar, Spherical, Multiplanar.}
	\label{fig:Figure4}
\end{figure}

\par In Fig. 3(b), it can be found that the SnS phenomena includes the birth-death caused by occlusion, and the channel characteristics of multipaths will also change greatly along the elements in near-field region, which is the characteristic that is not depicted in the standard [17]. Therefore, the planar wave modeling can be used within the SI, while near-field effects can be considered between stationary intervals. The multiplanar model can be considered. In the 3GPP channel model [17], the array antenna can be viewed as a point source, generating the same cluster angle under the assumption of far-field planar waves. The near-field antenna array is divided into SIs, and the same cluster angle is generated within the SI, while the angle varies between the SIs, which can be generated by establishing statistical angle models under different scenarios. To generate the non-stationarity over the array, it is necessary to chose different parameters of the ray/cluster at the transmitter for each sub-array.

\section{Verification of Near-Field Channel Characteristics in the New Mid-band}

\subsection{Measurement Configuration and Environment}	

\par Fig. 5 shows a laboratory room as the measurement scenario. Combined with the high precision automatic mechanical sliding platform, the vertical polarized omnidirectional antenna is translated to form a large-scale virtual array with the near-field region. The vector network analyzer (VNA)-based channel measurement system can be utilized to sound the propagation channel. The scenarios are divided into LOS and OLOS environments. In the OLOS environment, the antenna array is partially obscured by a baffle. The equipment required for measurement and corresponding parameter settings are given in Table \uppercase\expandafter{\romannumeral1}.

\begin{table}[htbp]
	\centering
	\caption{Parameters of Massive MIMO Channel Sounding Platform}
	\setlength{\tabcolsep}{0.1 mm}
	\label{my-label}
	\renewcommand{\arraystretch}{1.3}
	\setlength{\tabcolsep}{10 mm}
	\begin{tabular}{c|c}
		\hline \hline
		Parameter    & Performance  \\ \hline
		Frequency [GHz]    & 11-15  \\ \hline
		Tx power [dBm]     & 10  \\  \hline
		Antenna  type & Omnidirectional  \\  \hline
		Antenna height [m] & 2.5 \\  \hline
		Polarization & Vertical   \\   \hline \hline
	\end{tabular}
\end{table}

\begin{figure}[htbp]
	\xdef\xfigwd{\columnwidth}
	\setlength{\abovecaptionskip}{0.1 cm}
	\centering
	\begin{tabular}{cc}
		\includegraphics[width=4.0cm,height=3.5cm]{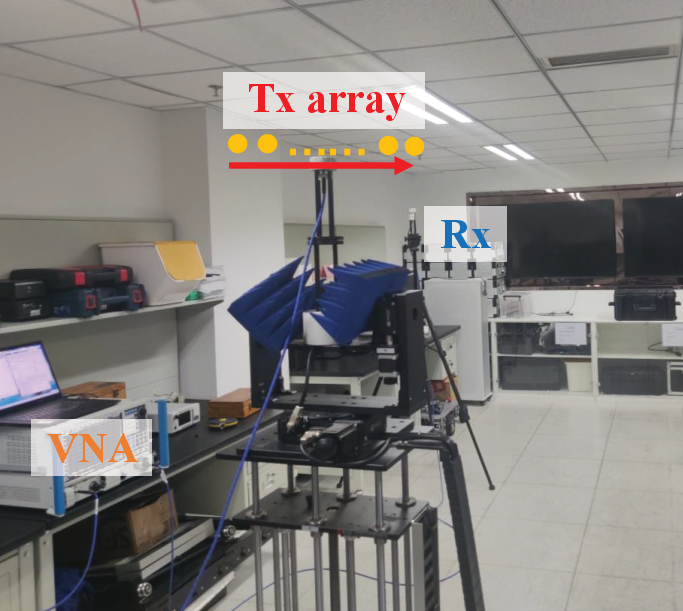} \hspace{-2mm}
		& \includegraphics[width=4.0cm,height=3.5cm]{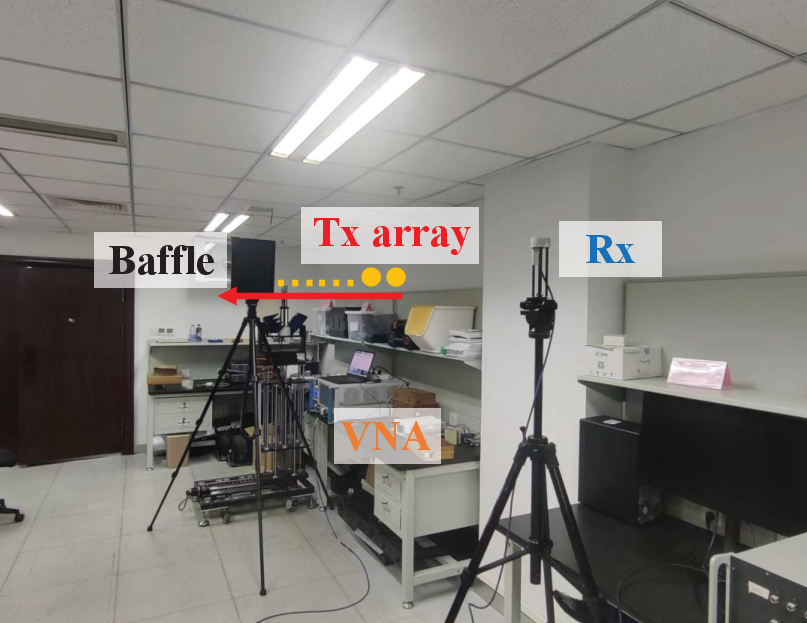}\\
		{\footnotesize\sf (a)} &	{\footnotesize\sf (b)} \\		
	\end{tabular}
	\caption{ Near-field channel measurement scenario. (a) LOS. (b) OLOS.}
	\label{fig:Figure5}
\end{figure}

\subsection{Near-Field Channel Characterization}

\subsubsection{Power Delay Spectrum}

\par In Fig. 6(a), the non-stationary properties caused by near-field are not obvious in LOS environment. In the OLOS environment shown in Fig. 6(b), the baffle has a significant effect. In the baffle range (starting with the 26th array element), the power can be seen to be significantly reduced in the blocked area. It can be observed that multi-path has a great influence on the main path, and the total power fluctuates slightly in LOS environment. However, in OLOS environment, occlusion still leads to the reduction of receiving power, and there are the significant SnS characteristics.

\begin{figure}[htbp]
	\xdef\xfigwd{\columnwidth}
	\setlength{\abovecaptionskip}{0.1 cm}
	\centering
	\begin{tabular}{cc}
		\includegraphics[width=4.2cm,height=3.5cm]{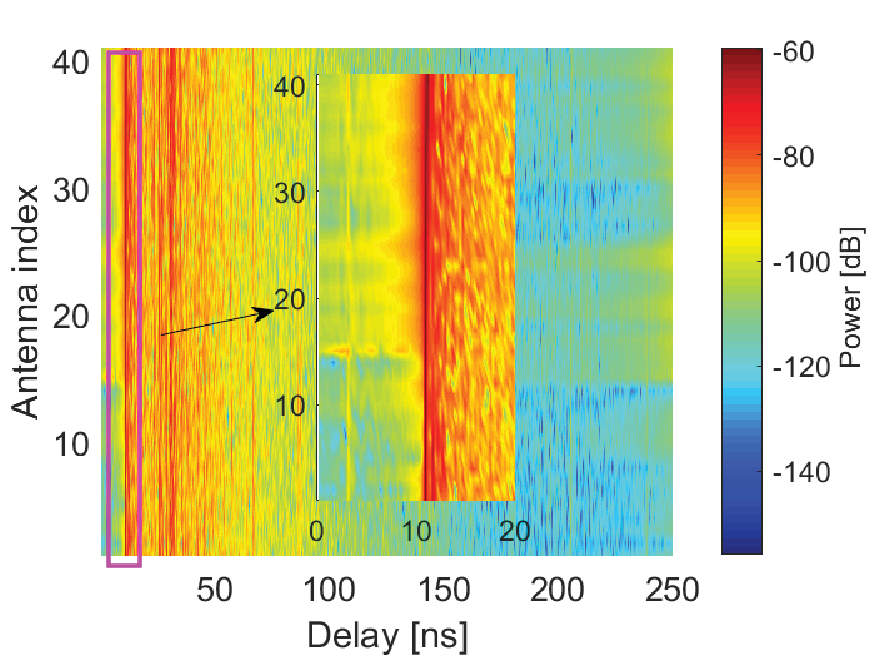} \hspace{-5mm}
		& \includegraphics[width=4.2cm,height=3.5cm]{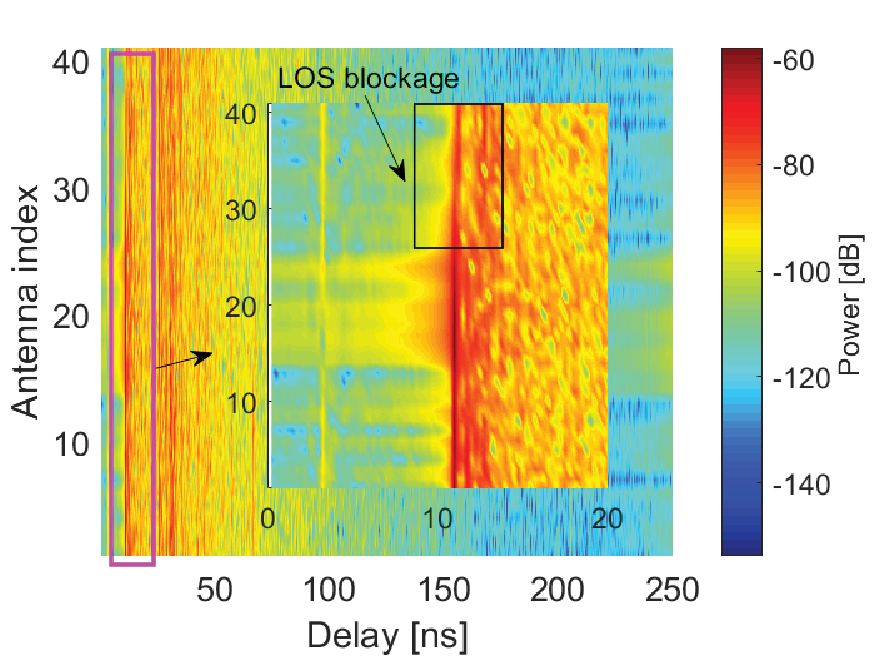}\\
		{\footnotesize\sf (a)} &	{\footnotesize\sf (b)} \\		
	\end{tabular}
	\caption{ Power on the array at the new mid-band. (a) LOS. (b) OLOS.}
	\label{fig:Figure6}
\end{figure}

\subsubsection{Delay Spread and Phase}

\par The RMS delay spreads and phase on the array at new mid-band in the LOS and OLOS environment are given in Fig. 7.



\begin{figure}[htbp]
	\xdef\xfigwd{\columnwidth}
	\setlength{\abovecaptionskip}{0.1 cm}
	\centering
	\begin{tabular}{cc}
		\includegraphics[width=4.2cm,height=3.5cm]{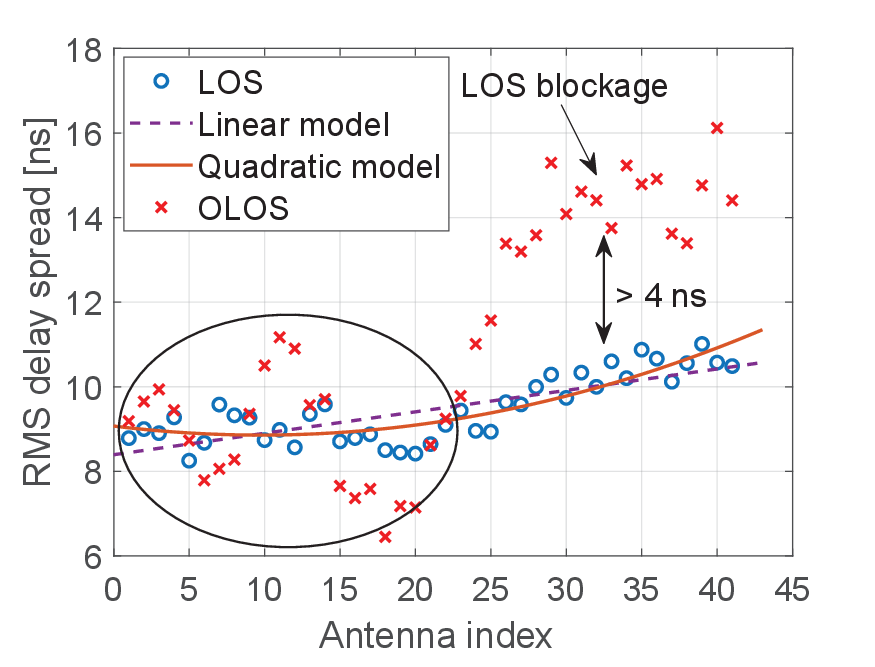} \hspace{-6mm}
		& \includegraphics[width=4.2cm,height=3.5cm]{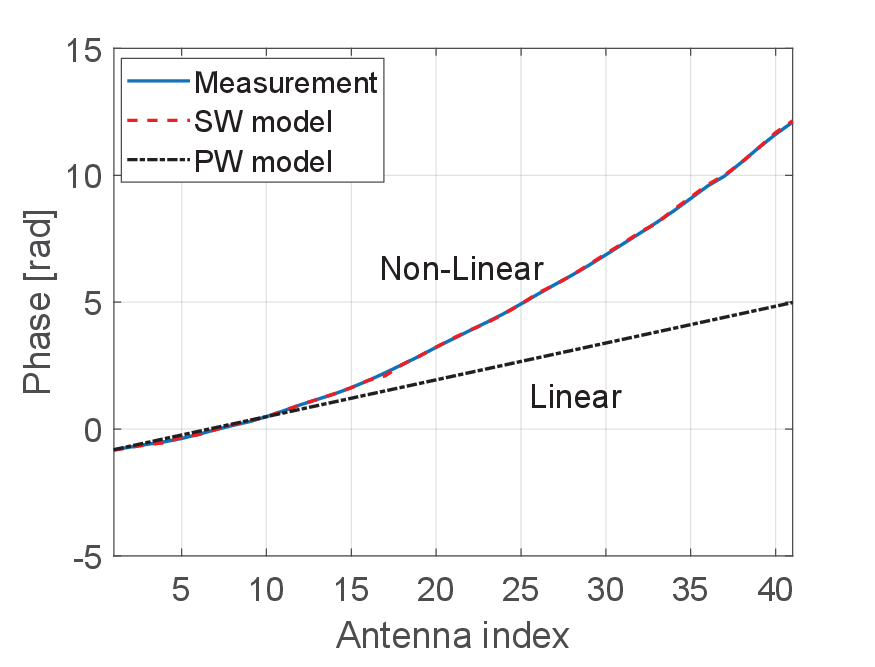}\\
		{\footnotesize\sf (a)} &	{\footnotesize\sf (b)} \\		
	\end{tabular}
	\caption{ Channel characteristics. (a) Delay spread. (b) Phase of LOS path.}
	\label{fig:Figure7}
\end{figure}


\par In the LOS environment, the fluctuation of delay spread is about 2 ns. The polynomial model can be used to generate the delay spread on the array, because the power changes in the array domain is quadratic. The delay spread can also be generated by using a first-order polynomial fitting model. In the OLOS environment, when part of the antenna array elements are completely blocked, the delay spread increases by about 5 ns, and the SnS phenomenon is significant. The results show that the SnS analysis is not only related to distance, but also the difference between LOS and NLOS scenarios is very obvious. In addition, by comparing Fig. 7(a), the delay spread also fluctuates slightly in the non-occluded part of the array (black ellipse), indicating that obstacles cause diffraction effects on this part of the channel, resulting in changes in channel state information. The results show that the influence of occlusions on SnS is very important for the study of near-field spatial nonstationarity. In Fig. 7(b), the phase obtained by measurement is in good agreement with the spherical-wave signal model.

\subsection{Stationary Interval}

\par To explore the influence of the environment on the SI on the array, Fig. 8 represents the correlation on the different array elements in the LOS/OLOS environments.

\begin{figure}[htbp]
	\xdef\xfigwd{\columnwidth}
	\setlength{\abovecaptionskip}{0.1 cm}
	\centering
	\begin{tabular}{cc}
		\includegraphics[width=4.2cm,height=3.5cm]{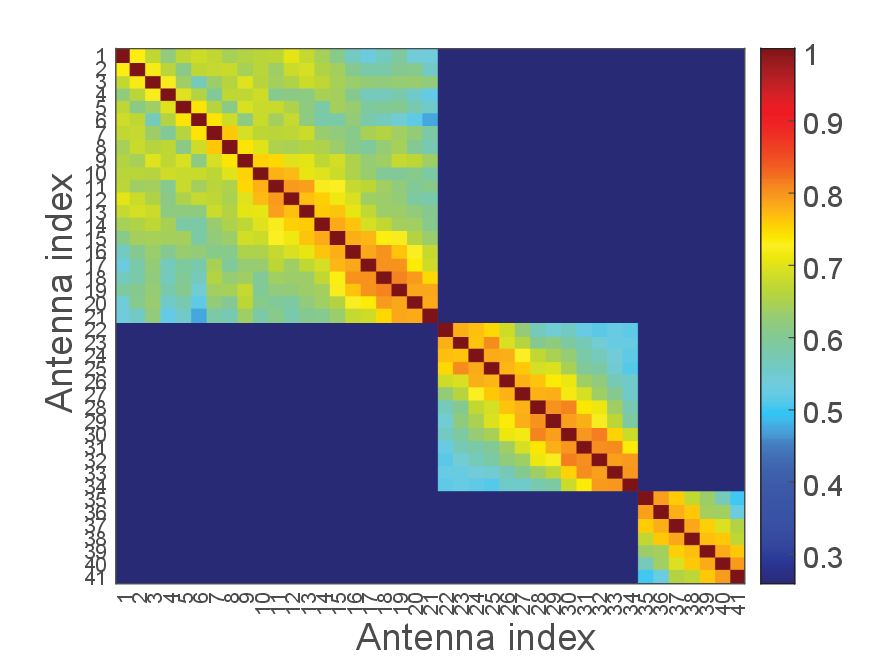} \hspace{-6mm}
		& \includegraphics[width=4.2cm,height=3.5cm]{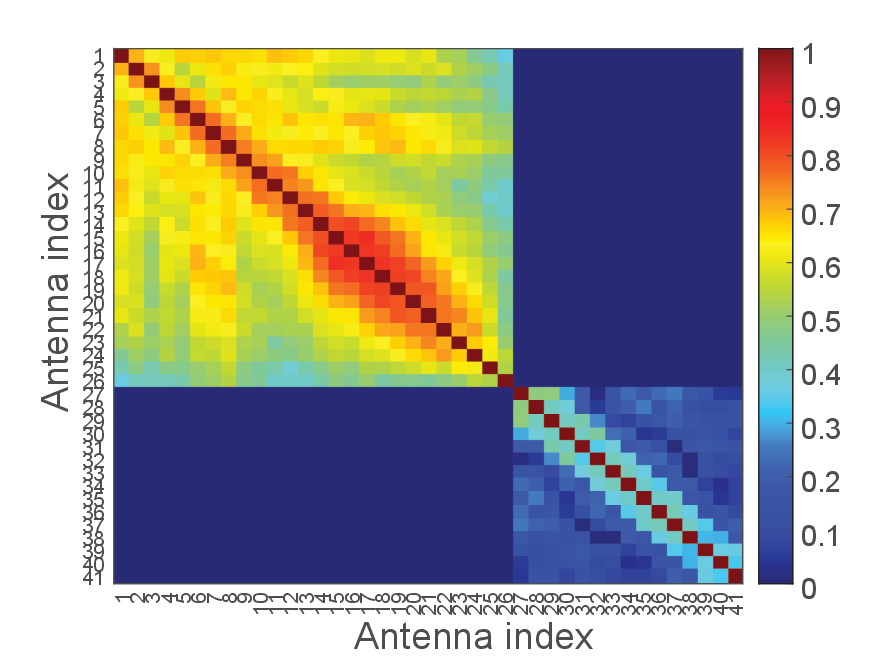}\\
		{\footnotesize\sf (a)} &	{\footnotesize\sf (b)} \\		
	\end{tabular}
	\caption{The SI on the array based on channel correlation. (a) LOS. (b) OLOS.}
	\label{fig:Figure8}
\end{figure}

\par Fig. 8 shows that the channel correlation matrix in LOS and OLOS environments has the very significant difference. The results show that obstacles can significantly change the SnS characteristics on the array. In Fig. 8(a), it can be seen that the SI is getting smaller along the array. The main reason is that the distance between Rx and each element on the Tx array gradually increases, and it can be observed the azimuth angular dispersion of the departure angle becoming more pronounced. The results show that the angle has a great influence on the spatial non-stationarity in the near-field region. In Fig. 8(b), it can be observed that when obstructions are present, more significant SnS characteristics will be caused. Therefore, the occlusion is one of the main factors leading to spatial non-stationarity in near-field channel modeling. Not only different types of scenarios, but also LOS/NLOS environments should explore the different recommended values of SI based on statistical data analysis.

\section{Conclusion}

\par In this paper, the channel measurements and characterization have been investigated for NFRC. The near-field spherical-wave signal model is derived and proposed. The boundary division methods of spherical wave and planar wave are introduced based on the SI. Based on the new mid-band near-field channel measurements, the near-field channel characteristics in LOS and OLOS environments are discussed, including delay domain and spatial domain. Compared with LOS environment, the occlusion will reduce the received power in OLOS environment, and there are the significant SnS characteristics.

\par For future work, it is necessary to further characterize the near-field channel model for 3GPP, including large and small scale channel characteristic parameters.

\ifCLASSOPTIONcaptionsoff
\newpage
\fi

\section*{Acknowledgment}

This work was supported in part by National Natural Science Foundation of China (62201086, 62101069, 92167202), National Science Fund for Distinguished Young Scholars (61925102), and BUPT-CMCC Joint Innovation Center.

\end{document}